\documentclass[sigconf]{acmart}
\usepackage{caption}
\usepackage{subcaption}
\usepackage[inline]{enumitem}
\usepackage{amsmath}
\usepackage{bm}
\usepackage{booktabs} 
\usepackage{multirow}
\usepackage{lipsum}
\usepackage{multicol}

\renewcommand{\baselinestretch}{0.97}

\AtBeginDocument{%
  \providecommand\BibTeX{{%
    \normalfont B\kern-0.5em{\scshape i\kern-0.25em b}\kern-0.8em\TeX}}}

\DeclareRobustCommand*{\ora}{\overrightarrow}

\def \cU {\mathcal{U}}
\def \cI {\mathcal{I}}
\def \cS {\mathcal{S}}

\copyrightyear{2020}
\acmYear{2020}
\setcopyright{acmcopyright}\acmConference[KDD '20]{Proceedings of the 26th ACM SIGKDD Conference on Knowledge Discovery and Data Mining}{August 23--27, 2020}{Virtual Event, CA, USA}
\acmBooktitle{Proceedings of the 26th ACM SIGKDD Conference on Knowledge Discovery and Data Mining (KDD '20), August 23--27, 2020, Virtual Event, CA, USA}
\acmPrice{15.00}
\acmDOI{10.1145/3394486.3403051}
\acmISBN{978-1-4503-7998-4/20/08}
\begin{document}
\fancyhead{}

\title{Directional Multivariate Ranking}

\author{Nan Wang}
\affiliation{
    \department{Department of Computer Science}
    \institution{University of Virginia} 
    \city{Charlottesville}
    \state{VA 22904}
    \country{USA}
 }
\email{nw6a@virginia.edu}

\author{Hongning Wang}
\affiliation{
    \department{Department of Computer Science}
    \institution{University of Virginia} 
    \city{Charlottesville}
    \state{VA 22904}
    \country{USA}
 }
\email{hw5x@virginia.edu}

\renewcommand{\shortauthors}{Nan Wang and Hongning Wang}

\begin{abstract}
User-provided multi-aspect evaluations manifest users' detailed feedback on the recommended items and enable fine-grained understanding of their preferences.
Extensive studies have shown that modeling such data  greatly improves the effectiveness and explainability of the recommendations. However, as ranking is essential in recommendation, there is no principled solution yet for collectively generating multiple item rankings over different aspects.

In this work, we propose a directional multi-aspect ranking criterion to enable a holistic ranking of items with respect to multiple aspects. Specifically, we view multi-aspect evaluation as an integral effort from a user that forms a vector of his/her preferences over aspects. Our key insight is that the direction of the difference vector between two multi-aspect preference vectors 
reveals the pairwise order of comparison. Hence, it is necessary for a multi-aspect ranking criterion to preserve the observed directions from such pairwise comparisons. We further derive a complete solution for the multi-aspect ranking problem based on a probabilistic multivariate tensor factorization model. Comprehensive experimental analysis on a large TripAdvisor multi-aspect rating dataset and a Yelp review text dataset confirms the effectiveness of our solution. 
\end{abstract}



\maketitle

\section{Introduction}
\label{sec:intro}


Users' multi-aspect evaluation of items provides more informative feedback about their preferences than merely an overall evaluation \cite{viewpoint, EFM, bptf}. In return, predicting multiple rankings with respect to different aspects elevates a user's freedom in interacting with the recommended items \cite{aspectrec}. For example, in hotel recommendation, a user may prefer to have the recommended hotels ranked by the aspect \textit{Value} that he/she cares mostly, while the other user may choose to rank by \textit{Service} or a simple combination of a few aspects. 

Users' evaluations on multiple aspects are dependent and compose a holistic assessment of an item \cite{aspectrec,MTER}. 
For instance, \textit{Value} and \textit{Service} might be correlated in a user's evaluations of hotels; but if we simply estimate two separate rankers for these two aspects, such correlation cannot be exploited or maintained.
This urges us to formulate multi-aspect ranking as a structured prediction problem across items and aspects.  
Most existing ranking solutions, such as Bayesian personalized ranking (BPR) \cite{BPR}, are only designed for single-aspect ranking. If applied, one has to isolate the aspects, thus fails to uphold the dependency among them. This inevitably leads to suboptimal ranking performance, such as inconsistent rankings between correlated aspects. 
Besides, we should note the multi-aspect ranking problem is not simply a multi-task learning problem \cite{RMTL, MTER}, in which one learns a shared set of parameters that govern the rankings in multiple aspects. As such a solution decomposes the multi-aspect ranking problem into single aspect ranking sub-tasks and simply joins them via parameter sharing, there is no guarantee that the ranking on different aspects will correlate with each other. 

\begin{figure}[!h]
    \centering
    \includegraphics[width=0.96\linewidth]{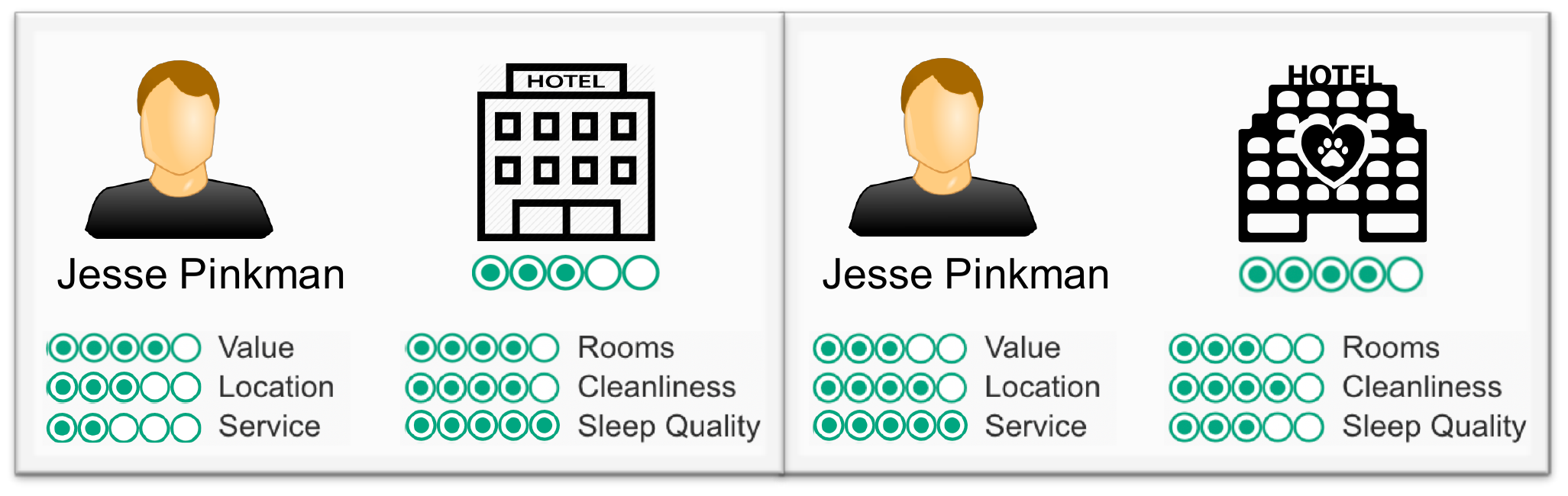}
    \caption{Multi-aspect evaluations on two hotels from the same user on TripAdvisor.}
    \label{fig:review}
    \vspace{-2mm}
\end{figure}

The fundamental research question in multi-aspect recommendation, i.e., \textit{how to generate multiple rankings with respect to different aspects collectively}, remains under-explored. The challenge lies in constructing the pairwise comparison between a pair of items, which should maintain structured dependency among aspects.
In single aspect ranking, the pairwise order can be directly obtained by the result of binary comparisons \cite{BPR}, e.g., a five-star item is better than a three-star one. Nevertheless, in multi-aspect scenario, the comparison between two items is \textit{multifaceted}, where the results of cross aspect comparisons carries users' implicit emphasis over aspects. 
For instance, Figure \ref{fig:review} shows a pair of multi-aspect evaluations from the same user on two hotels selected from TripAdvisor. 
Although the first hotel receives a higher rating on the aspect \textit{Value}, it gets a lower rating on \textit{Service}. Moreover, the difference on \textit{Value} is one star, but it is a three-star difference on \textit{Service}, which reflects the relative significance of the comparison on these two aspects. 
Intuitively, a mis-order when ranking items on an aspect with a larger observed pairwise difference should incur a higher ranking loss than that on an aspect with a smaller difference. 
Therefore, the pairwise comparison between two items in the multi-aspect case should follow two important criteria:
\begin{enumerate*}[label=(\roman*)]
  \item preserve single-aspect order on each aspect;
  \item maintain relative significance of the comparison on each aspect.
\end{enumerate*}

To provide a unified criterion for multi-aspect ranking, we take a vectorized representation of users' multi-aspect evaluation on each item, which enables explicit modeling of the dependency across aspects. Under this representation, the single-aspect orders as well as relative significance of comparison across aspects naturally embed in the \emph{direction} of the difference vector between two items' rating vectors. Specifically, the sign of each element in the difference vector suggests the pairwise order of the compared items on the corresponding single aspect; and each element's \emph{relative} value indicates the significance of the pairwise comparison on this aspect. Furthermore, both the sign and the \emph{relative} value can be reliably captured in the direction of the difference vector, agnostic to the magnitude, which leads to the pairwise \emph{directional order} defined by the direction of the difference vector between two items.  
Hence, to rank items with respect to multi-aspect comparisons, one should look for methods that maintain the directions of difference vectors. This gives rise to our directional analysis based ranking criterion: a multi-aspect ranking criterion should maximize the likelihood of the observed directions of the pairwise difference vectors. 
It thus motivates us to explore directional statistics  \cite{Bahlmann:2006:DFO:1220964.1221165}, which impose probability distributions for modeling the difference vector as a circular variable \cite{SphereDispersion}. 
Nonetheless, existing directional statistics are often limited to normalized magnitude and are not able to model the dependency among different dimensions. 

In this work, we devise new directional statistics to realize our ranking criterion. Specifically, for a given ranking model, we compute likelihood of the observed directions by taking \emph{line integral} along the directions over the model's imposed distribution on the difference vectors. An estimate of the underlying model is thus required to maximize this likelihood. We refer to this new ranking criterion as Directional Multi-aspect Ranking. 
We demonstrate the application of this criterion with a probabilistic multivariate tensor factorization model (PMTF), which imposes a multivariate Gaussian distribution over users' multi-aspect evaluations. 
Although we focus on PMTF to complete the solution, the proposed criterion is generally applicable to a broader scope of ranking models, as long as they define a probability distribution of the pairwise difference vector, such as deep neural network based models \cite{NCF}.

To evaluate the multi-aspect ranking performance of the proposed solution, we collected a large multi-aspect rating dataset from TripAdvisor. As reviews are common and abundant aspect-level evaluations can be extracted \cite{EFM,lara}, we also utilize a Yelp review dataset to verify our method's generalization to the review data. Extensive experiment comparisons against several baselines demonstrate the effectiveness of our solution.

\section{Related Works}
\label{sec:related}
Recommender systems rely on users' feedback to learn their preferences. Such preferences are oftentimes multifaceted. For example, a user's evaluations on different aspects of a hotel, such as \textit{Cleanliness}, \textit{Location}, and etc., provide her detailed aspect-level preferences \cite{lara,wang2011latent}. 
Empirical results show that such multi-aspect evaluations contribute to an in-depth comprehension of users' preferences and help improve the quality of recommendations \cite{DBLP:journals/corr/abs-1210-3926,jmars, bptf, HFT,EFM,MTER}. 

Recommendation is essentially a ranking problem; and various ranking criteria, such as Bayesian Personalized Ranking (BPR) \cite{BPR}, have been proposed to directly optimize the ranking quality. However, existing ranking solutions only address single-aspect ranking, and thus are usually only applied to optimize the overall ranking \cite{MTER, fact}. 
One direction in learning to rank studies the problem of document ranking with multi-aspect relevance labels \cite{l2rma}, via label aggregation or model aggregation. However, such approaches only obtain an aggregated function to generate a final overall ranking of documents, rather than to jointly learn a set of ranking functions for each aspect. To the best of our knowledge, there is no principled solution yet for ranking with multiple aspects collectively, though it has great practical value in modern recommender systems. 

Multi-aspect ranking cannot be simply addressed by multi-task learning \cite{RMTL, MTL} over a set of single aspect ranking tasks, because simply sharing parameters cannot capture the correlation or consistency of the generated rankings among different aspects. The improved performance from multi-task learning in modeling multi-aspect data \cite{wang2019bpmr, MTER} mostly comes from their refined user and item representations, by sharing observations to conquer data sparsity. Similarly, multi-objective optimization \cite{wei-MOO} is not a suitable solution. Methods in this family are mostly designed to handle trade-offs between two or more conflicting objectives. But in multi-aspect ranking problem, the learning of rankings under different aspects should not conflict with each other; instead, they should mutually support each other via the intrinsic dependency relation. 

There are several recent studies predicting multi-aspect ratings. For example, \citet{CCC} proposed to construct the list of aspects as a chain and predict a user's rating on each aspect one by one. 
\citet{ExSAE} extends the stacked auto-encoders with modified input layer and loss function to enable the learning of multi-aspect ratings. But such solutions cannot directly optimize item ranking resulted from the predicted ratings, and therefore there is no guarantee for their multi-aspect recommendation quality.

One of the main challenges in the task of multi-aspect ranking is to define the order between two preference vectors. 
This leads to solutions rooted in directional statistics \cite{Bahlmann:2006:DFO:1220964.1221165}, e.g., modeling the difference vector as a circular random variable \cite{batmanghelich-etal-2016-nonparametric,SphereDispersion}. However, such solutions restrict the vectors on a unit sphere and cannot model the dependency among different dimensions. Navarro et al. \cite{mgvm} generalize the von Mises distribution to a multivariate setting to capture the correlations among different dimensions. But due to the lack of a closed-form analytic expression, its complicated approximations prevent efficient parameter estimation for practical use. We derive our new directional statistics via line integral over the distributions on the difference vectors specified by a given probabilistic ranking model. This results in the likelihood that should be maximized by a learnt ranking model. 
\section{Methodology}
\label{sec:method}
In this section, we elaborate our proposed solution for multi-aspect ranking. We first formally define the pairwise directional order between two preference vectors. Based on it, we propose a general multi-aspect ranking criterion in Section \ref{dmrc}. In Section \ref{dmr}, we implement the criterion via a generalized probabilistic multivariate tensor factorization method, where we complete the solution by deriving the explicit form of the directional multivariate ranking objective function under the factorization model.

In our subsequent discussions, we assume there are $M$ users, $N$ items and $K$ aspects, and user preferences on those aspects are indicated by explicit numerical ratings. The proposed solution can be easily extended to implicit feedback with binary evaluations. For each user, the rated items are assumed to be preferred over unrated items \cite{BPR, NCF}. Without loss of generality, we consider the \textit{Overall} rating as one of the aspects, which evaluates an item in general. 
Let $\cU$ be the set of all users and $\cI$ the set of all items. For each multi-aspect rating evaluation, we denote $\bm R_{ui} \in \mathbb{R}^K$ as a random \textit{row} vector; and the observed rating vector $\bm r_{ui} = (r_{ui1}, r_{ui2}, ..., r_{uiK})$ is an observation of $\bm R_{ui}$ from user $u$ to item $i$. 

\subsection{Directional Order for Multi-aspect Ranking}
\label{dmrc}

\begin{figure}
  \centering
  \includegraphics[width=\linewidth]{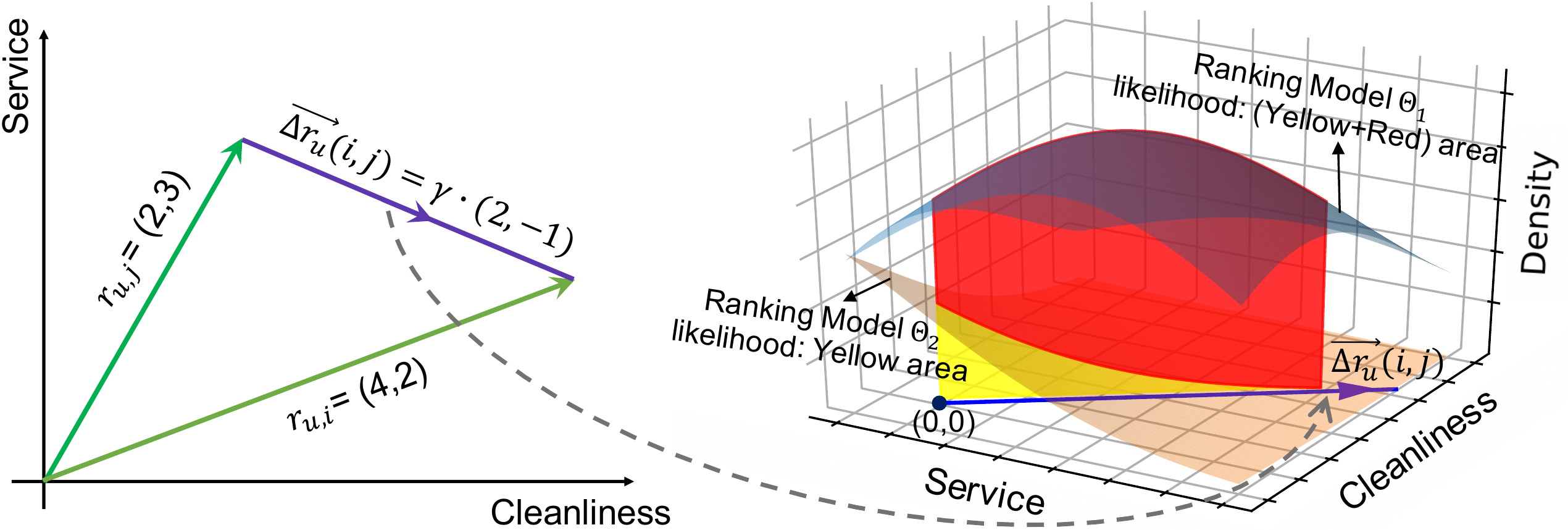}
  \caption{Illustration of the directional multi-aspect ranking criterion. The left side shows the observed direction from $\bm r_{u,i}$ to $\bm r_{u,j}$, denoted as $\ora{\Delta r_u}(i,j) = \{ \gamma\cdot(\bm r_{ui}-\bm r_{uj})|\,\gamma\in\mathbb{R}^+\}$. The right side illustrates the likelihood obtained from line integral along $\ora{\Delta r_u}(i,j)$ over the distribution of $\bm D_{uij} = (\bm R_{ui}-\bm R_{ui})$ imposed by two possible ranking models.}
  \label{fig:directional}
  \vspace{-1em}
\end{figure}

\subsubsection{Comparison on Multi-aspect Evaluations}
There is clear evidence that users' evaluations across multiple aspects of items are not independent \cite{Sahoo2008OnMR}. Hence, the revealed preference order between two items cannot be trivially expressed as one is better than the other. Instead, complex comparison between a pair of evaluations exists, e.g., a user may give hotel A a higher rating on the aspect of \textit{Cleanliness} than hotel B, but a lower rating on the aspect of \textit{Service}. 

The challenge is then to define the pairwise order between two rating vectors. Specifically, we require the order to encode both pairwise comparisons on each aspect and the dependency and relative significance of comparison across aspects.
This leads to our proposal of defining such a pairwise order from the comparison between two rating vectors as the \emph{direction} of the difference vector $\bm r_{ui} - \bm r_{uj}$. As illustrated in Figure \ref{fig:directional} (left) in a two-dimensional rating space, there are two aspect ratings in each evaluation, one for the aspect \textit{Cleanliness} and the other for \textit{Service}. Given two rating vectors $(4,2)$ and $(2,3)$, the difference vector $(2,-1)$ encodes the comparison between $\bm r_{ui}$ to $\bm r_{uj}$. Specifically, the single-aspect preference orders are indicated by the sign in each dimension of the difference vector, and the relative significance of each aspect-level comparison is reflected in the absolute values in each dimension. In this example, user $u$ prefers \textit{Cleanliness} of hotel $i$ over hotel $j$ with a 2-star difference, while believes \textit{Service} in hotel $i$ is slightly worse than that of hotel $j$ with a 1-star difference. Thus the comparison on \textit{Cleanliness} is more significant than the comparison on \textit{Service}. 

As a ranking problem, the exact magnitude of the difference vector is not essential, which will only impose unnecessary constraints for optimization \cite{BPR}. As long as the direction of the difference vector is maintained in the prediction, the single-aspect orders are still preserved by the sign of each corresponding dimension. 
In particular, for any point along this direction, the relative comparison is maintained by the ratio across the dimensions. Therefore, we denote the direction of the difference vector from $\bm r_{uj}$ to $\bm r_{ui}$ as a \textit{ray}: $\overrightarrow{\Delta r_u}(i,j) = \{ \gamma\cdot(\bm r_{ui}-\bm r_{uj})|\,\gamma\in\mathbb{R}^+\}$, which encodes the revealed pairwise comparison result across the aspects over these two items. As the scalar $\gamma$ is only required to be positive, it relaxes the comparison's dependency on the magnitude of the difference vector, while preserving the preference order under each aspect and the relative comparison significance across aspects. 


\subsubsection{Directional Multi-aspect Ranking Criterion}
With the directional orders encoded via rays in the multi-aspect space, we can now formulate the multi-aspect ranking criterion
as maximizing the likelihood of the observed pairwise directions. 
We first create the set of pairs $\cS := \{(u,i,j)|u\in \cU; i, j\in \cI \wedge i\neq j\}$. We use $>_u$ to denote the desired multi-aspect ranking orders under user $u$. The items are assumed to be independently evaluated
\cite{BPR}. Hence, the likelihood of the observed pairwise orders in a collection of users specified by a given ranking model $\Theta$ can be written as:
\begin{equation}
    \mathcal{L}(\Theta) = \prod\limits_{u\in \cU}p(>_u|\Theta) 
    = \prod\limits_{(u,i,j)\in \cS}p(\overrightarrow{\Delta r_u}(i,j)|\Theta)
    \label{eq:direct_ll}
\end{equation}
According to the definition of $\overrightarrow{\Delta r_u}(i,j)$, all points on the ray $\overrightarrow{\Delta r_u}(i,j)$ with $\gamma\in(0,\infty)$ are valid for representing the same pairwise comparison. Thus the likelihood of the observed direction for each pair is proportional to the line integral \cite{Tang2007Mathematical} along the ray over the density of the difference vector imposed by the ranking model $\Theta$:
\begin{align*}
    &p(\overrightarrow{\Delta r_u}(i,j)|\Theta) \\
    &\propto \int_{0}^{\infty} p\Big((\bm R_{ui}-\bm R_{uj}) = \gamma\cdot(\bm r_{ui}-\bm r_{uj})|\Theta\Big)\times\lVert\bm r_{ui}-\bm r_{uj}\rVert d\gamma \nonumber
\end{align*}
where $\lVert\bm x\rVert$ is the $L_2$ norm of a vector. For convenience, we denote $\bm D_{uij} = (\bm R_{ui}-\bm R_{uj})$ and $\bm d_{uij} = (\bm r_{ui}-\bm r_{uj})$. Since both rating vectors are given as observed, $\lVert\bm r_{ui}-\bm r_{uj}\rVert$ is a constant. We can safely remove this term when we compute the log-likelihood for optimization in later derivations, and from now on we will not include it to simplify our notations. The resulting likelihood function in Eq~\eqref{eq:direct_ll} is then reformulated as:
\begin{equation}
\label{eq:lineint}
    \mathcal{L}(\Theta) = \prod\limits_{(u,i,j)\in \cS}\int_{0}^{\infty} p(\bm D_{uij} = \gamma\cdot\bm d_{uij}|\Theta) d\gamma
\end{equation}

Figure \ref{fig:directional} (right) illustrates how we derive the likelihood of observed pairs and optimize the ranking model. Follow the two-aspect case in Figure \ref{fig:directional} (left), we first need to construct the density function of the pairwise difference vector $\bm D_{uij}$, represented by the two surfaces in the figure imposed by two models $\Theta_1$ and $\Theta_2$. Then we take line integral along the observed direction $\overrightarrow{\Delta r_u}(i,j)$ for this pair with respect to the density functions, as the shaded areas indicate, to get the likelihood. As a result, the likelihood of the observed direction from $\Theta_1$ is the yellow plus red area, while the likelihood from $\Theta_2$ is the yellow area. For optimization, our criterion will chose the ranking model which specifies the density that maximizes the likelihood, i.e., choose $\Theta_1$ over $\Theta_2$. 

From a Bayesian perspective, our goal of finding the optimal parameter set $\Theta$ is to maximize the posterior:
\begin{equation*}
    p(\Theta|>_u, u\in \cU) \propto \mathcal{L}(\Theta)\,p(\Theta)
\end{equation*}
where $p(\Theta)$ specifies a prior distribution of the ranking model. However, note that in Eq.~\eqref{eq:lineint}, zero is included in the integral for the scalar $\gamma$. This can possibly lead to a trivial ranking model that places the mean vector of the predicted difference distribution very close to zero. This cannot provide meaningful and robust pairwise comparisons.
To prevent the learnt distribution from concentrating on zero differences, we add a small margin $\xi$ constraint to $\gamma$ in the integral, i.e., $\gamma\in(\xi,\infty)$ and $\xi>0$, which leads to our directional multi-aspect ranking criterion:
\begin{align}
\label{eq:dmr}
    & \max\limits_\Theta \,\ln p(\Theta|>_u, u\in \cU) \\ \nonumber
         =& \max\limits_\Theta \,\ln \mathcal{L}(\Theta)\,p(\Theta) \\ \nonumber
         =& \max\limits_\Theta \sum\limits_{(u,i,j)\in \cS}\ln \int_{\xi}^{\infty} p(\bm D_{uij} = \gamma\cdot\bm d_{uij}|\Theta) d\gamma + \ln p(\Theta)
\end{align}
for obtaining a ranking model on multiple aspects. 

With the directional orders, the derived criterion provides a principled way to obtain an optimal model for ranking multiple aspects jointly. This criterion views the ranking on multiple aspects as a structured prediction task, which directly models dependency among aspects. Nevertheless, in order to apply the criterion to multi-aspect ranking, we need the ranking model $\Theta$ to specify a valid distribution of $\bm D_{uij}$ in the multi-aspect space. In the next section, we will derive our probabilistic multivariate modeling of $\bm D_{uij}$, which imposes a probability distribution over the directions of pairwise comparions and captures the dependency among aspects. It thus completes our solution to the multi-aspect ranking problem. 

\subsection{A Complete Directional Multivariate Ranking Solution}
\label{dmr}
\subsubsection{Probabilistic Multivariate Tensor Factorization}
The multi-aspect ranking criterion proposed in Section \ref{dmrc} is generally applicable to any valid probabilistic model of multi-aspect evaluations. In this work, we adopt probabilistic tensor factorization (PTF) for its popularity and flexibility in modeling the generation of multi-aspect ratings. PTF has demonstrated the promising performance in learning the latent representations from multi-aspect data \cite{bptf, temporal}. 

In particular, three matrices $\bm U\in\mathbb R_+^{M\times d}$, $\bm V\in\mathbb R_+^{N\times d}$, $\bm W\in\mathbb R_+^{K\times d}$ are used to represent latent factors of users, items and aspects, respectively, with each row vector representing a single entity. The predicted multi-aspect rating vectors are then obtained as $\hat{\bm{r}}_{ui} = ( U_u * V_i)\ W^\top$, where $*$ denotes the element-wise product. One issue with conventional probabilistic tensor factorization is the default assumption that the ratings on multiple aspects are independently sampled from $K$ Gaussian distributions:
\begin{equation*}
    p(\bm r_{ui}|\bm U,\bm V,\bm W, \sigma) = \prod\limits_{k=1}^{K} \mathcal{N}(r_{uik} | <U_u,V_i,W_k>, \sigma)
\end{equation*}
where $\mathcal{N}(\mu,\sigma)$ is a Gaussian distribution with mean $\mu$ and standard deviation $\sigma$. This directly prevents us from modeling and discovering the dependency among aspects. To capture the dependency among different aspects, we consider the rating vector that user $u$ gives item $i$ as sampled from a \emph{multivariate} Gaussian distribution:
\begin{equation*}
    \bm R_{ui} \sim \mathcal{N}\big((U_u*V_i)W^\top, \bm\Sigma_{ui}\big)
\end{equation*}
where $(U_u*V_i)W^\top$ is the resulting mean vector of dimension $K$, and $\bm\Sigma_{ui}$ is a $K\times K$ covariance matrix for a user-item pair. 

The emphasis on different aspects may vary from different users; and the same logic applies to items. Therefore, the covariance matrix 
should be personalized for users and items to capture individualized preferences. To formally quantify the personalized covariance composition for the observation of a user-item pair, we take a perspective stemmed from the Item Response Theory (IRT) \cite{Fayers2004Item}. IRT states that the probability of a specific response is determined by the item's attribute and the subject's individuality \cite{Lin}. Mapping it back to our problem, we assume that the covariance matrix of an observed rating vector is a result from two separated channels of preference emphasis from the corresponding user and item. Specifically, we assign each user $u$ and each item $i$ a personalized covariance matrix $\bm\Sigma_u^U$ and $\bm\Sigma_i^V$, respectively. The covariance matrix for this pair of user-item is given by: $\bm\Sigma_{ui} = \lambda\bm\Sigma_u^U + (1-\lambda)\bm\Sigma_i^V$,
where $\lambda \in [0,1]$ is a hyper-parameter, reflecting the relative strength of influence from the user side in this composition. 
The covariance matrices should also be learned from observations. 
A by-product of this design is that the learnt covariance matrices can serve as the basis for personalized aspect-level explanations \cite{EFM, MTER}, e.g., when the recommendation list is ranked by the overall ratings, the most correlated aspects should be selected for explanation. 

We name this factorization solution as Probabilistic Multivariate Tensor Factorization (PMTF). It enables the modeling and ranking of multiple aspects collectively in a multivariate fashion. Although we use the multivariate Gaussian distribution to model the dependency among aspects, other kinds of multivariate distributions from the exponential family, such as multivariate Poisson distribution, can also be applied to characterize the generation of multi-aspect rating vectors. Later we will demonstrate how this family of probability distribution ensures a closed form solution of likelihood.  

\subsubsection{Multi-aspect Ranking with PMTF}
Recall that in Eq.~\eqref{eq:dmr}, what we need is the estimated density on difference vector $\bm D_{uij} = \bm R_{ui} - \bm R_{uj}$. Under PMTF, 
the density function of $\bm D_{uij}$ is obtained as:
\begin{equation}
\label{eq:multi-normal}
    p(\bm D_{uij} = \gamma\cdot\bm d_{uij} |\Theta) = \mathcal{N}\Big(\gamma\cdot\bm d_{uij} |\, \big(U_u*(V_i-V_j)\big)W^\top, \bm\Sigma_{uij}\Big)
\end{equation}
where $\bm\Sigma_{uij} = \bm\Sigma_{ui} + \bm\Sigma_{uj}$ as the imposed density rating vectors are independent and follow multivariate Gaussian distribution. 
Applying the density function of the difference vector in Eq.~\eqref{eq:multi-normal} to the likelihood function defined on the observed direction in Eq.~\eqref{eq:direct_ll} with a margin $\xi$, we can obtain an explicit expression of $\mathcal{L}(\Theta)$ as:
\small
\begin{align}
\label{eq:mulvarrank_ll}
    \mathcal{L}(\Theta) = \prod\limits_{(u,i,j)\in \cS}&\int_{\xi}^{\infty} p(\bm D_{uij} = \gamma\cdot\bm d_{uij}|\Theta) d\gamma \\\nonumber
    = \prod\limits_{(u,i,j)\in \cS}&\int_{\xi}^{\infty}\mathcal{N}\Big(\gamma\cdot\bm d_{uij} |\, \big[U_u*(V_i-V_j)\big]W^\top, \bm\Sigma_{uij}\Big) d\gamma \\\nonumber
    = \prod\limits_{(u,i,j)\in \cS}&\frac{1}{\sqrt{(2\pi)^K|\bm\Sigma_{uij}|}}\exp\Big(-\frac{1}{2}(C-\frac{B^2}{A})\Big)\\\nonumber
    &\times\sqrt{\frac{\pi}{2A}}\Big(1-erf\big(\sqrt{\frac{A}{2}}(\xi-\frac{B}{A})\big)\Big)\\\nonumber
\end{align}
\normalsize
where $A = \bm d_{uij}\bm\Sigma_{uij}^{-1}\bm d_{uij}^\top,\, B = \bm d_{uij}\bm\Sigma_{uij}^{-1}\hat{\bm d}_{uij}^\top,\, C = \hat{\bm d}_{uij}\bm\Sigma_{uij}^{-1}\hat{\bm d}_{uij}^\top$; $\hat{\bm d}_{u,i,j} = \big(U_u*(V_i-V_j)\big)W^\top$ is the estimated mean from the resulting tensor; $|\bm X|$ is the determinant of matrix $\bm X$; and $erf(x)=\frac{2}{\sqrt\pi}\int_{0}^{x}e^{-t^2}dt$ is the Gauss error function. The detailed derivations of this likelihood function 
can be found in the Appendix. 

By treating the latent factors in PMTF as random variables, we place zero-mean spherical Gaussian priors on the latent factor matrices $\bm U,\bm V,\bm W$ \cite{pmf}:
\begin{equation*}
p(\bm U|\sigma_U^2) = \prod\limits_{u=1}^M \mathcal{N}(U_u|\bm 0,\sigma_U^2 \bm I),
\end{equation*}
$p(\bm V|\sigma_V^2)$ and $p(\bm W|\sigma_W^2)$ are given similarly. 

The drawback in modeling personalized covariance matrices is that we need sufficient observations on each user and item. For the users and items with only a handful of observations, it is infeasible to estimate accurate covariance matrices from data. To address the data sparsity issue for such users and items, we introduce the normal-inverse-Wishart distribution as the global prior for each personalized covariance matrix, which is defined on real-valued positive (semi-)definite matrices \cite{Conjugate, wishart}:
\begin{equation*}
    p(\bm\Sigma|\Psi,\nu) = \frac{|\Psi|^{\nu/2}}{2^{\nu K/2}\Gamma_K(\frac{\nu}{2})}|\bm \Sigma|^{-(\nu + K + 1)/2}e^{\frac{1}{2}tr(\Psi\bm\Sigma^{-1})}
\end{equation*}
where $tr(\bm X)$ is the trace of matrix $\bm X$. $\nu$ and $\Psi$ are hyper-parameters, which impose that the \emph{priori} covariance matrix is from $\nu$ observations with sum of pairwise deviation products $\Psi$, i.e., $\Psi = \nu\bm\Sigma_{priori}$. In practice, we can set $\Sigma_{priori}$ to the estimated global covariance matrix from the observed data across all users and items. In this way, updating the personalized covariance matrices under this prior is equivalent to modifying the global covariance matrix for each individual with local observations. 

With the likelihood function $\mathcal{L}(\Theta)$ and the priors $p(\Theta)$ given above, we can now substitute the corresponding components in Eq.~\eqref{eq:dmr} to get the explicit objective function under the PMTF model for personalized directional multivariate ranking:
\begin{align}
\label{eq:mulvarrank_obj}
    &\max\limits_\Theta \,\ln p(\Theta|>_u) \\ \nonumber
    = &\max\limits_\Theta\sum\limits_{(u,i,j)\in \cS} \Big[-\frac{1}{2}\big(\ln|\bm\Sigma_{uij}|+ C - \frac{B^2}{A} + \ln(2A)\big) + \ln\big(1 - erf(Z)\big)\Big] \\ \nonumber
    +&\sum\limits_{u=1}^M \Big[-\frac{\nu+K+1}{2}\ln|\bm\Sigma^U_u| + \frac{1}{2}tr(\Psi{\bm\Sigma_u^{U}}^{-1}) + \frac{\nu}{2}\ln|\Psi| - \ln 2^{\nu K/2}\Gamma_K(\frac{\nu}{2})\Big]\\\nonumber 
    +&\sum\limits_{i=1}^N \Big[-\frac{\nu+K+1}{2}\ln|\bm\Sigma^V_i| + \frac{1}{2}tr(\Psi{\bm\Sigma_i^{V}}^{-1}) + \frac{\nu}{2}\ln|\Psi| - \ln 2^{\nu K/2}\Gamma_K(\frac{\nu}{2})\Big]\\\nonumber 
    +&\ln p(U|\sigma_U^2) + \ln p(V|\sigma_V^2) + \ln p(W|\sigma_W^2) + const \nonumber
\end{align}
where $Z=\sqrt{\frac{A}{2}}\big(\xi-\frac{B}{A}\big)$. Though the error function could not be expressed by elementary functions, the derivative of it can be explicitly calculated by $\frac{d}{dx}erf(x)=\frac{2}{\sqrt\pi}e^{-x^2}$. Therefore, analytic solutions exist for gradient-based optimization for Eq. \eqref{eq:mulvarrank_obj}. Optimizing the above objective function gives us the parameters for the PMTF model under the directional multi-aspect ranking criterion.

\subsubsection{Model Optimization}
To efficiently estimate both the latent factors and personalized covariance matrices, we appeal to mini-batch stochastic gradient descent (SGD) for optimization. In each iteration, we have two steps for parameter update. In the first step, we sample a batch of rating vector pairs from $\cS$, and update the latent factors in $\bm U, \bm V, \bm W$ with SGD. In the second step, we sample another batch of pairs and update the associated covariance matrices. We employ the adaptive gradient descent method \cite{adagrad} for better convergence, which dynamically incorporates the updating trace to perform more informative and faster gradient-based learning. 

For optimizing a multi-aspect ranking model like PMTF, we need the training triples of $(u,i,j)$ in the order of $O(M\times N\times N)$, which makes it crucial to find an efficient sampling strategy for gradient descent. Following \cite{BPR}, we adopt the method of bootstrap sampling with replacement, which samples triples in each iteration uniformly at random. Due to the transitivity of pairwise orders, sampling a subset of all possible pairs is sufficient to determine the full ranking. 
The time complexity of optimization scales linearly with the number of users, items, aspects, and the dimension of latent factors. We terminate the optimization procedure when the maximum iteration is reached or it converges on the hold-out validation set. 

Finally yet importantly, in order to learn valid covariance matrices, we need to ensure they are symmetric and positive semi-definite \cite{covariance}. To satisfy this constraint, we set each covariance matrix in a form of $\bm\Sigma = \bm L\bm L^\top$, where $\bm L$ is an arbitrary matrix of the same dimension with $\bm\Sigma$. Then for any non-zero column vector $\bm x$, we have $\bm x^\top\bm L\bm L^\top\bm x = (\bm L^\top\bm x)^\top(\bm L^\top\bm x) \geq 0$, which ensures the positive semi-definite constraint. Also, the symmetric property is naturally satisfied as $\bm L\bm L^\top = (\bm L\bm L^\top)^\top$. With this decomposition of covariance matrix, we update the matrices $\bm L$ in the optimization procedure and re-construct the covariance matrices with the resulting $\bm L$. The gradient for all parameters are provided in the Appendix.

\section{Experiments}
\label{sec:exp}
In this section, we evaluate the proposed solution for multi-aspect ranking. We collected a large multi-aspect rating dataset from TripAdvisor. To generalize the solution to the feature-level evaluations in review text data, we also extract a multi-aspect evaluation dataset with phrase-level sentiment analysis \cite{EFM} from Yelp reviews. Extensive experimental analysis from different perspectives are performed. Considerable improvements over various kinds of baselines confirm the effectiveness of our multi-aspect ranking solution.

\subsection{Experiment Setup} 
\noindent\textbf{$\bullet$ Dataset.} We use the TripAdvisor multi-aspect rating dataset and the Yelp review dataset for evaluation. The pre-processing procedures for each of them are described below. The statistics of the resulting datasets are summarized in Table \ref{tab:stas}. If not noted specifically, each rating/observation refers to a multi-aspect rating vector with possible missing values. \\
\textbf{TripAdvisor Multi-aspect Ratings.} The data was collected from May 2014 to September 2014 from  TripAdvisor\footnote{Data was collected before they forbid practitioners to crawl their data.}. There are over 3 million multi-aspect ratings from over 1 million users on around 18 thousand hotels. Each multi-aspect rating vector consists of one overall rating plus seven specific aspect ratings: [`\textit{Overall}', `\textit{Sleep Quality}', `\textit{Service}', `\textit{Value}', `\textit{Rooms}', `\textit{Cleanliness}', `\textit{Location}', `\textit{Check in / front desk}']. 
All ratings are integers ranging from 1 to 5. By filtering out users and hotels who have less than 5 observations, we obtain the final dataset for evaluation, consisting of over 1M multi-aspect rating vectors from around 100k users and 17k hotels.

\begin{table}[t]
\centering
\caption{Statistics of the two datasets. The last column means the number of multi-aspect rating vectors.}
\vspace{-2mm}
\label{tab:stas}
\begin{tabular}{@{}ccccc@{}}
\toprule
Dataset     & \#user  & \#item & \#aspect & \#ratings\\ \midrule
TripAdvisor & 100,005 & 17,257 & 8        & 1,057,217 \\
Yelp        & 52,050  & 34,729 & 14       & 704,825   \\ \bottomrule
\end{tabular}
\vspace{-1em}
\end{table}

\noindent\textbf{Yelp Reviews.} The review data is provided by Yelp challenge\footnote{https://www.yelp.com/dataset/challenge}, which contains user reviews mostly on restaurants. In order to obtain explicit aspect-level evaluations, we apply phrase-level sentiment analysis \cite{EFM,sentianaly} to extract the aspects and sentiment polarities for each aspect from review content. The sentiment score for each aspect is then mapped to the range of 1-5 as explicit aspect-level ratings. We refer the readers to \cite{EFM, sentianaly} for the detailed procedure. We only maintain the aspects that appear more than 200k times in the entire review corpus, which give us multi-aspect rating vectors consisted of one overall rating plus thirteen aspect ratings: [`\textit{Overall}', `\textit{Food}', `\textit{Service}', `\textit{Bar}', `\textit{Meal}', `\textit{Chicken}', `\textit{Menu}', `\textit{Price}', `\textit{Staff}', `\textit{Restaurant}', `\textit{Sauce}', `\textit{Cheese}', `\textit{Taste}', `\textit{Pizza}']. By filtering out users and restaurant with less than 5 reviews, we obtain a dataset of 52k users, 34k restaurants and 704k multi-aspect ratings. 

\begin{table*}[]
\centering
\caption{Multi-aspect ranking performance with MAP and NDCG@K. The percentage of improvement is from the best over the second best method on each dataset.}
\label{tab:ndcg}
\begin{tabular}{ccccccccc}
\hline
TripAdvisor & \multicolumn{8}{c}{By Overall Aspect} \\ \hline
 & PTF & EFM & MTER & M-BPR & Ex-SAE & SPS-MTL & DMR & \multicolumn{1}{l}{Improvement} \\
MAP & 0.0828 & 0.1007 & 0.2146 & 0.2179 & 0.1944 & 0.2206 & \textbf{0.2397} & 8.66\%* \\
NDCG@10 & 0.1004 & 0.1246 & 0.2461 & 0.2472 & 0.2109 & 0.2512 & \textbf{0.2775} & 10.47\%* \\
NDCG@50 & 0.2083 & 0.2259 & 0.3576 & 0.3532 & 0.3196 & 0.3587 & \textbf{0.3781} & 5.41\%* \\ \hline
TripAdvisor & \multicolumn{8}{c}{By Average on All Aspects} \\ \hline
 & PTF & EFM & MTER & M-BPR & Ex-SAE & SPS-MTL & DMR & Improvement \\
MAP & 0.0816 & 0.0994 & 0.2044 & 0.2118 & 0.1897 & 0.2156 & \textbf{0.2298} & 6.59\%* \\
NDCG@10 & 0.0984 & 0.1198 & 0.2302 & 0.2416 & 0.2084 & 0.2466 & \textbf{0.2687} & 8.96\%* \\
NDCG@50 & 0.1906 & 0.2138 & 0.3419 & 0.3459 & 0.3116 & 0.3508 & \textbf{0.3695} & 5.33\%* \\ \hline
Yelp & \multicolumn{8}{c}{By Overall Aspect} \\ \hline
 & PTF & EFM & MTER & M-BPR & Ex-SAE & SPS-MTL & DMR & Improvement \\
MAP & 0.1208 & 0.1389 & 0.3492 & 0.3566 & 0.3318 & 0.3695 & \textbf{0.4089} & 10.66\%* \\
NDCG@10 & 0.1635 & 0.1896 & 0.3907 & 0.3991 & 0.3685 & 0.4036 & \textbf{0.4558} & 12.93\%* \\
NDCG@50 & 0.3051 & 0.3268 & 0.4793 & 0.4814 & 0.4521 & 0.4892 & \textbf{0.5314} & 8.63\%* \\ \hline
Yelp & \multicolumn{8}{c}{By Average on All Aspects} \\ \hline
 & PTF & EFM & MTER & M-BPR & Ex-SAE & SPS-MTL & DMR & Improvement \\
MAP & 0.1035 & 0.1217 & 0.2874 & 0.2931 & 0.2674 & 0.3086 & \textbf{0.3312} & 7.32\%* \\
NDCG@10 & 0.1375 & 0.1691 & 0.3426 & 0.3509 & 0.3176 & 0.3588 & \textbf{0.3953} & 10.17\%* \\
NDCG@50 & 0.2547 & 0.2735 & 0.4205 & 0.4296 & 0.3959 & 0.4376 & \textbf{0.4696} & 7.31\%* \\ \hline
\end{tabular}
\\\emph{*p}-value < 0.05 under paired t-test.
\end{table*}

\noindent\textbf{$\bullet$ Baselines.}
We compare with various kinds of baselines, ranging from traditional matrix/tensor factorization models to multi-task learning based solutions, and the recent neural networks based ranking methods, to provide a comprehensive evaluation of our proposed solution for multi-aspect ranking. 
Below we briefly summarize the details about the baselines. \\    
\textbf{PTF}: Probabilistic Tensor Factorization \cite{bptf,Rai:2015:SPT:2832747.2832775}. It is a traditional tensor factorization method based on CP decomposition. It treats multiple aspects independently, and it is optimized to minimize the square loss on each aspect, i.e., a rating regression model.  \\
\textbf{M-BPR:} Multi-task Bayesian Personalized Ranking \cite{BPR}. BPR is a generic single-aspect ranking criterion that aims to maximize the likelihood of the observed pair-wise preference order. 
Specifically, we extend it to the multi-aspect case by  multi-task learning. The ranking on each aspect is considered as a learning task, and the user and item representations are shared among all tasks to couple the learning tasks. The objective functions for all tasks are added together for optimization. We also tried other forms to integrate the loss from individual tasks, e.g., taking a product; but summation gives us the best performance in our evaluation.\\
\textbf{EFM}: Explicit Factor Models \cite{EFM}. It is a joint matrix factorization algorithm that models a user's aspect-level attention as well as an item's aspect-level quality measured by the opinion ratings. The aspect-level preference predictions are given by the product of corresponding predicted user-aspect attention and item-aspect quality.\\
\textbf{MTER}: Explainable Recommendation via Multi-task Learning \cite{MTER}. It is a joint tensor factorization algorithm that models user's detailed opinions on individual aspects. BPR is imposed in it, but only on the overall rating. As we do not consider users' opinionated review content in our evaluation, we exclude its content modeling component and only maintain the rating tensor in MTER.\\
\textbf{SPS-MTL}: Soft Parameter Sharing for Multi-task Learning in deep neural networks \cite{MTL}. We adopt a five-layer MLP for each single-aspect rating prediction task with soft parameter sharing for multi-task learning. The distance between the parameters of the first three hidden layers is regularized in order to encourage the parameters to be similar. Its inputs are the user and item pairs and outputs are the predicted aspect ratings for each aspect. The same pair-wise logistic loss as in \cite{BPR} is adopted for model training on each aspect. \\
\textbf{Ex-SAE}: Extended Stacked Auto Encoders \cite{ExSAE}. It extends SAE by incorporating an extra layer to cater the requirement of multi-aspect rating prediction. We set the input and output to the multi-aspect rating vectors, with 5 hidden layers for both the encoder and decoder. 

We refer to our proposed directional multivariate ranking solution as DMR. Both datasets are split by 70\% for training, 15\% for validation, and 15\% for testing. We used grid search to find the optimal hyper-parameters on the validation set. The datasets and implementations can be found at: \url{https://github.com/MyTHWN/Directional-Multivariate-Ranking}.

\subsection{Results on Multi-aspect Top-$K$ Ranking}
We start our experimental analysis with top-$K$ multi-aspect ranking. As most previous solutions make recommendations by the \textit{Overall} rating aspect, we first report MAP and NDCG@$\{10,50\}$ obtained from ranking by the predicted \textit{Overall} rating. This will validate the benefit of jointly optimizing the rankings on multiple aspects by exploiting their correlations. Meanwhile, we also report the average MAP and NDCG@$\{10,50\}$ over all aspects to study how the multi-aspect rankings mutually improve each other. 
Specifically in DMR, we set $\lambda$ to 0.5, the margin $\xi$ to 0.2 on TripAdvisor and 0 on Yelp data, the latent factor dimension to 10, and initial learning rate to 0.03. $\bm\Sigma_{priori}$ is estimated from the observations of corresponding dataset across all users and items. The maximum number of iterations is set to 40,000 with 2,000 multi-aspect rating pairs in each iteration.   

The results are shown in Table \ref{tab:ndcg}. 
Paired t-test is performed with \textit{p-value} $< 0.05$ to confirm the significance of the improvement from DMR against the baselines. In general, directly optimizing the ranking performance (e.g., MTER, M-BPR, SPS-MTL, DMR) can achieve much better top-$K$ ranking performance than point-wise predictions (e.g., PTF, EFM, Ex-SAE). By modeling the dependency and joint optimization in DMR, its ranking performance on all aspects is further boosted. Comparing MTER with M-BPR and SPS-MTL, although the ranking loss is imposed on all aspects in M-BPR and SPS-MTL, while it is only on the overall rating in MTER, their overall ranking performance is very close to each other. This indicates that separating the multi-aspect ranking task into individual single-aspect objectives is insufficient to exploit their dependency nor to mutually improve the ranking performance on all aspects. The improvement from DMR demonstrates the advantage of taking a holistic view of the multi-aspect evaluation. 

\begin{figure}
  \centering
  \begin{subfigure}[b]{0.236\textwidth}
    \centering
    \includegraphics[width=\textwidth]{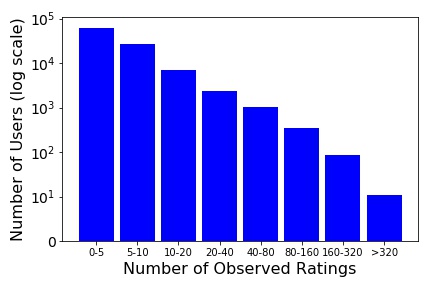}
    \caption{Distribution of users.}
    \label{fig:user_dist}
  \end{subfigure}
  \hfill
  \begin{subfigure}[b]{0.236\textwidth}
    \centering
    \includegraphics[width=\textwidth]{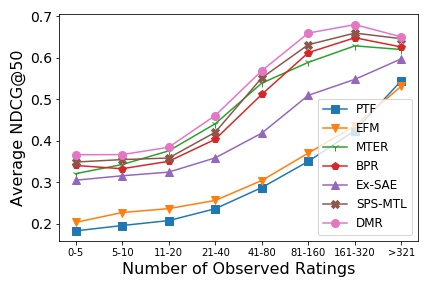}
    \caption{NDCG@50 in user groups.}
    \label{fig:ndcg_usernum}
  \end{subfigure}
  \caption{NDCG@50 over all aspects across user groups defined by the number of training instances in each user.}
  \label{fig:user_num_ndcg}
\vspace{-1em}
\end{figure}

To better understand the ranking performance under different kinds of users, we group users based on the number of observations they have in the training set on TripAdvisor and investigate the ranking performance in each user groups. Figure \ref{fig:user_num_ndcg} demonstrates the distribution of users according to their observed rating vectors in training set and the corresponding average NDCG@50 over all aspects in each user group. The results on Yelp show similar patterns; but due to space limit, we omit them in this section. We can find that the distribution of users with respect to their number of observations is highly skewed: Over 80\% users have less than 10 rating vectors. Figure \ref{fig:user_num_ndcg} (b) shows DMR consistently outperformed all baselines in different user groups, even with only a handful of ratings for training. This advantage is further amplified when there are more observations in heavy users. The reason is that in DMR, we take multi-aspect ranking as an integrated learning task and explicitly model the dependency among aspects. It enables the learning of latent factors to fully exploit the correlations and consistency among multi-aspect ratings and thus mutually enhance the ranking performance on all aspects even with sparse data. 

\subsection{Pairwise Order Accuracy with Different Confidence Levels}
\begin{table}[]
\centering
\caption{Pair-wise order accuracy across all aspects.}
\label{tab:pair_accu}
\begin{tabular}{@{}ccc@{}}
\toprule
Accuracy & TripAdvisor & Yelp \\ \midrule
PTF & 0.6295 & 0.7252 \\
EFM & 0.6812 & 0.7951 \\
MTER & 0.8256 & 0.8964 \\
M-BPR & 0.8139 & 0.8978 \\
Ex-SAE & 0.7926 & 0.8695 \\
SPS-MTL & 0.8204 & 0.9027 \\
DMR-I & 0.8292 & 0.9254 \\
DMR & \textbf{0.8384} & \textbf{0.9325} \\ \bottomrule
\end{tabular}
\vspace{-3mm}
\end{table}
Besides receiving a ranked list of items, oftentimes a user needs to compare different aspects on two recommended items. Thus we also evaluated the pairwise order accuracy, i.e., whether a model can correctly differentiate the relative comparison of two items on each aspect. We randomly sample 150k rating vector pairs from the testing set of TripAdvisor and 100k pairs from the testing set of Yelp. We define accuracy as the proportion of correctly predicted orders on all aspects in all sampled pairs:
\begin{equation*}
    accuracy = \frac{\sum\limits_{(u,i,j)}\sum\limits_{k=1}^K \bm{1}\big[(r_{uik}-r_{ujk})(\hat r_{uik}-\hat r_{ujk})>0\big]}{\sum\limits_{(u,i,j)}\sum\limits_{k=1}^K \bm{1}\big[r_{uik}-r_{ujk}\neq 0\big]}
\end{equation*}
where $\bm{1}[\cdot]$ is the indicator function. The pairwise accuracy over all aspects is reported in Table \ref{tab:pair_accu}. In the table, we introduce a new baseline DMR-I to investigate the benefits of the personalized covariance modeling. DMR-I is a variant of DMR by replacing its personalized covariance matrices with a fixed \emph{identity} covariance matrix. Thus, DMR-I only updates the latent factors. The results in Table \ref{tab:pair_accu} demonstrate that DMR still provides the most accurate pairwise comparisons against all baselines, and imposing personalized correlations leads to better learned latent factors that correctly differentiate the comparisons.

\begin{figure}
  \centering
  \begin{subfigure}[b]{0.236\textwidth}
    \centering
    \includegraphics[width=\textwidth]{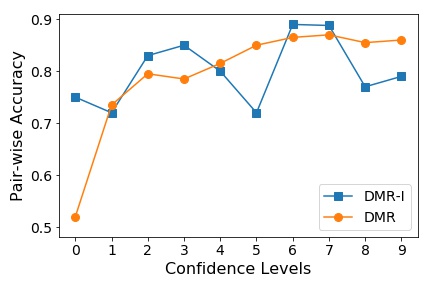}
    \caption{TripAdvisor}
    \label{fig:accu_trip}
  \end{subfigure}
  \hfill
  \begin{subfigure}[b]{0.236\textwidth}
    \centering
    \includegraphics[width=\textwidth]{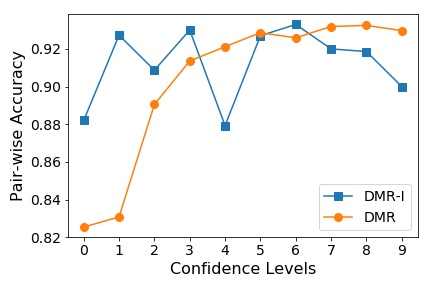}
    \caption{Yelp}
    \label{fig:accu_yelp}
  \end{subfigure}
  \vspace{-4mm}
  \caption{The pair-wise accuracy according to the confidence levels in DMR and DMR-I. The confidence levels are measured by the relative likelihood of the predicted pair-wise direction by Eq~\eqref{eq:mulvarrank_ll} from low to high.}
  \label{fig:pari_accuracy}
  \vspace{-2mm}
\end{figure}

So far, all the results are from the predicted mean rating vector $\hat{\bm r}_{ui}= (U_u*V_i)W^\top$, i.e., only using the latent factors to predict the ranking or pairwise orders.
However, we are not only estimating the mean, but fitting the exact distribution as shown in Figure \ref{fig:directional} (right) as well. A well estimated distribution should have its probability mass well align with the ground-truth direction of the difference vectors. As we use Gaussian distribution, the density of the estimated distribution is guided by the covariance matrix, which measures correlation among aspects. In order to investigate the quality of the learned covariance matrices, we evaluate the confidence of the predicted pairwise order.  
Intuitively, if the distribution is more \textit{concentrated} to the predicted direction, we should be more confident that it is close to the ground-truth direction. Therefore, for each selected pair, we take the line integral along the predicted direction as its confidence measure about the predicted order. Thus if the line integral along the predicted direction is high, it suggests a high confidence in the prediction. We divide the sampled pairs to 10 groups based on the confidence of the prediction from low to high, and calculate pairwise accuracy under each confidence level. As only DMR models the covariance across aspects, we compare it with its variant DMR-I in this experiment and report the results on both datasets in Figure \ref{fig:pari_accuracy}. 

With personalized covariance modeling in DMR, the accuracy evidently increases as we increase level of confidence. But for DMR-I, which fixes the covariance matrices to an identity matrix, there is no clear relation between accuracy and confidence levels. This experiment confirms the necessity and quality of learned personalized covariance matrices: It provides a means to assess the model's prediction confidence. A unique benefit is that informing users about the model's prediction confidence can also potentially increase users' trust on the rankings and recommendations.

\subsection{Effect of the Margin $\xi$}
\begin{figure}
  \centering
  \begin{subfigure}[b]{0.236\textwidth}
    \centering
    \includegraphics[width=\textwidth]{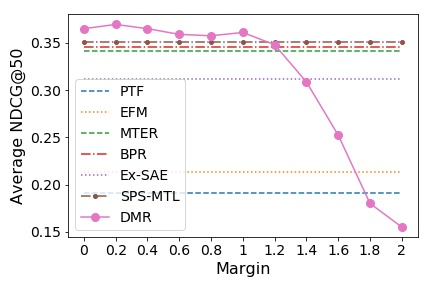}
    \caption{NDCG@50 against margin $\xi$.}
    \label{fig:margin_trip}
  \end{subfigure}
  \hfill
  \begin{subfigure}[b]{0.236\textwidth}
    \centering
    \includegraphics[width=\textwidth]{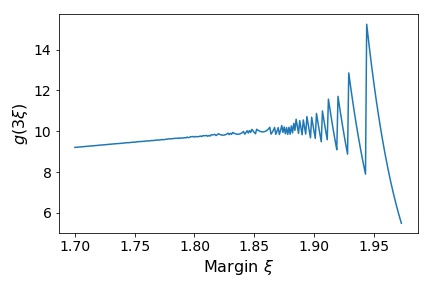}
    \caption{Gradient explosion against $\xi$.}
    \label{fig:explosion}
  \end{subfigure}
  \caption{Average NDCG@50 over all aspects with respect to margin $\xi$ on TripAdvisor; and the explosion of gradient when $\xi$ is too large.}
  \label{fig:margin}
\end{figure}

In order to avoid the learnt mean vectors concentrating on zero vectors, we impose a margin constraint when computing the likelihood via line integral in Eq~\eqref{eq:dmr}. However, during hyper-parameter tuning, we found that although a small margin can help achieve better rankings, the top-$K$ ranking performance drops dramatically when setting a too large margin $\xi$. This effect is shown in Figure \ref{fig:margin} (a) on the TripAdvisor dataset; and similar effect is also observed on the Yelp dataset. This might be a bit counter-intuitive at the first glance, as a larger margin should give us a stronger constraint on the predicted orders and better improved rankings. But this observation is expected for the following reasons. First, more pairwise orders will conflict with each other when we set a larger margin. Because a large margin tends to push the difference vectors to large magnitudes, i.e., orders with highly significant comparisons, which makes it more difficult to maintain all the input pairwise orders and creates many sub-optimal solutions. Second, more seriously, there is a numerical issue introduced by the sufficient statistics in the derived gradients of DMR's objective function. The gradients with respect to the latent factors in Eq.~\eqref{eq:mulvarrank_obj} have a common term $g(x) = e^{-x^2}/\big(1-erf(x)\big)$, where $x = c_{uij}\cdot\xi$ linearly increases with $\xi$; $c_{uij}$ increases at a constant rate that depends on the sampled user-item pair. When $x$ is large, $g(x)$ becomes unstable and explode numerically. We illustrate the explosion of gradient when $c_{uij}=3$ in Figure \ref{fig:explosion}. To avoid the issue, we suggest to set a relatively small margin in practice (usually smaller than 0.5).

\subsection{Learned Correlations for Explanation}
The learned personalized correlations contain a user's (item's) intrinsic emphasis on different aspects, and can be used to generate correlation-based explanations. For example, suppose the items are ranked for a user $u$ based on the predicted \textit{Overall} ratings. Then for each recommended item $i$, we can construct the correlation matrix from $\bm\Sigma_{ui}$ and use the most correlated aspects to the \textit{Overall} aspect as the personalized aspect-level explanation. The selected aspects (e.g., \textit{Cleanliness} and \textit{Room}) can be embedded to an explanation template to synthesize natural language explanations like: ``We recommend this hotel \textit{[A]} to you, because you may like its aspects of \textit{[Cleanliness]} and \textit{[Room]}.'' \cite{EFM, MTER}.

\begin{table}[!h]
\caption{Average rating difference between overall and the selected aspect (smaller is better).}
\label{tab:aver_diff}
\begin{tabular}{@{}ccccc@{}}
\toprule
Average Rating Difference & Random & MTER   & EFM    & DMR             \\ \midrule
TripAdvisor               & 1.3478 & 1.4572 & 0.9574 & \textbf{0.6962} \\
Yelp                      & 3.2471 & 4.1852 & 2.6916 & \textbf{2.2502} \\ \bottomrule
\end{tabular}
\end{table}

To evaluate the quality of the generated explanations with selected aspects, for each observed user rating vector in the testing set, we calculate the rating difference between the \emph{ground-truth} \textit{Overall} rating and the \emph{ground-truth} selected aspect's rating. In DMR, the aspect that has the highest predicted personalized correlation with the \textit{Overall} aspect is selected. If the selected aspect's rating is indeed highly dependent on the \textit{Overall} rating, we should expect a smaller \emph{ground-truth} rating difference between them. 
Based on this idea, we compare the average rating difference across all rating vectors from different selection strategies. EFM \cite{EFM} suggests to select the aspect based on a linear combination of a user's aspect attention and an item's aspect quality. MTER \cite{MTER} chooses the aspect with the highest predicted rating. In other baselines, as there is no explicit rule for selecting aspects for explanation, we use a uniformly random selection from the aspects as a baseline. The results are reported in Table \ref{tab:aver_diff}. From the results, correlation based aspect selection from DMR achieves significantly better performance in this evaluation of explainability. 
Compared with the performance on TripAdvisor dataset, the average rating difference on Yelp dataset is relatively higher. This is because the aspect ratings are extracted from reviews in this dataset, and there are lots of missing aspect ratings (treated as 0). But if a user did not mention an aspect in his/her review, it strongly indicate the aspect is not important for him/her. Therefore, the results on Yelp dataset arguably better emphasize a model's ability in recognizing important aspects in each user-item pair. 
This study demonstrates the benefit of modeling correlations for generating aspect-level explanations and it is obviously a better way than heuristics in other methods.

\section{Conclusions and Future Work}
\label{sec:conclusion}
In this paper, we propose a directional multivariate ranking solution for improving the learning of ranking functions on multiple aspects collectively. Instead of treating aspects in isolation, we view the multi-aspect ranking as an integrated task by defining the directional orders between two preference vectors for optimization. Comprehensive experimental results on two benchmark datasets prove the effectiveness of the solution. 

Our work aims to improve the rankings on multiple aspects jointly. A parallel problem is to provide a single ranking that can satisfy the preference orders of multiple aspects. As we have already defined the pairwise orders in the multi-aspect case, and the DMR criterion provides an integrated objective to maintain all single-aspect orders, the future goal is to find a way for generating a single ranking that follows the criterion. Besides, in our work, observed items in a user are assumed to be preferred over unobserved ones. However, such unobserved \emph{irrelevant} evaluations could never be verified, and therefore introduce unnecessary bias in model training. This potentially degenerates the rankers' performance. Some recent work has realized this issue and proposed counter-factual inference based solutions to reduce the bias \cite{IPS} in single aspect rating settings. But how to differentiate the missing evaluations to infer the preferences in multi-aspect setting is still largely unknown. We plan to further extend our solution to this setting, e.g., with partially or completely missing aspect ratings. 

\begin{acks}
We thank the anonymous reviewers for their insightful comments. This paper is based upon work supported by the National Science Foundation under grant IIS-1553568 and SCH-1838615.
\end{acks}

\bibliographystyle{ACM-Reference-Format}
\bibliography{reference}
\newpage
\section*{Appendix}
\subsection*{Derivation of Eq.~\eqref{eq:mulvarrank_ll}}
In order to get the explicit expression of Eq.~\eqref{eq:mulvarrank_ll}, we can first write out the probability of $p(\bm D_{uij} = \gamma\cdot\bm d_{uij}|\Theta)$ with the multivariate Gaussian distribution as follows: 
\begin{flalign}
\label{eq:gauss}
    &\mathcal{L}(\Theta)\\\nonumber
    =& \prod\limits_{(u,i,j)\in S}\int_{\xi}^{\infty} p(\bm D_{uij} = \gamma\cdot\bm d_{uij}|\Theta) d\gamma \\\nonumber
    =& \prod\limits_{(u,i,j)\in S}\int_{\xi}^{\infty}\mathcal{N}(\gamma\cdot\bm d_{uij} |\, [U_u*(V_i-V_j)]W^\top, \bm\Sigma_{uij}) d\gamma \\\nonumber
    =& \prod\limits_{(u,i,j)\in S}\frac{1}{\sqrt{(2\pi)^K|\bm\Sigma_{uij}|}}\int_{\xi}^{\infty}  exp\big(-\frac{1}{2}(\gamma\cdot\bm d_{uij}-[U_u*(V_i-V_j)]W^\top)\\\nonumber
    & \;\bm\Sigma_{uij}^{-1}(\gamma\cdot\bm d_{uij}-[U_u*(V_i-V_j)]W^\top)^\top\big) d\gamma
\end{flalign}
For simplicity of the notations, we can let:
\begin{align*}
&\hat{\bm d}_{u,i,j} = \big(U_u*(V_i-V_j)\big)W^\top;\;\\
&A = \bm d_{uij}\bm\Sigma_{uij}^{-1}\bm d_{uij}^\top,\; B = \bm d_{uij}\bm\Sigma_{uij}^{-1}\hat{\bm d}_{uij}^\top,\; C = \hat{\bm d}_{uij}\bm\Sigma_{uij}^{-1}\hat{\bm d}_{uij}^\top
\end{align*}
Then our goal is to calculate the integral in Eq.~\eqref{eq:gauss}. After extracting the terms not included in the integral and take a square formula with respect to $\gamma$, we have:
\begin{flalign}
\label{eq:integral}
    &\int_{\xi}^{\infty}exp\big(-\frac{1}{2}(\gamma\cdot\bm d_{uij}-\hat{\bm d}_{u,i,j})\bm\Sigma_{uij}^{-1}(\gamma\cdot\bm d_{uij}-\hat{\bm d}_{u,i,j})^\top\big) d\gamma &\\\nonumber
    &=\; exp(-\frac{1}{2}C)\cdot\int_{\xi}^{\infty} exp\big(-\frac{1}{2}(\gamma^2A - 2\gamma B)\big) d\gamma & \\\nonumber
    &=\; exp(-\frac{1}{2}C)\cdot\int_{\xi}^{\infty} exp\big(-\frac{1}{2}A(\gamma - \frac{B}{A})^2 + \frac{B^2}{2A}\big) d\gamma & \\\nonumber
    &=\; exp(\frac{B^2}{2A}-\frac{1}{2}C)\cdot\int_{\xi}^{\infty} exp\big(-\frac{1}{2}A(\gamma - \frac{B}{A})^2\big) d\gamma &
\end{flalign}
Based on the definition of the error function and the integral of the exponential function, we can easily get a general equation: 
\begin{align*}
\int_{a}^{b}exp(-x^2)dx = \frac{\sqrt\pi}{2}\big(erf(b)-erf(a)\big)
\end{align*}
which can be applied to get the solution of the integral in Eq.~\eqref{eq:integral} as:
\begin{align}
\label{eq:errorfunc}
    &\int_{\xi}^{\infty}exp\big(-\frac{1}{2}A(\gamma - \frac{B}{A})^2\big) d\gamma \\\nonumber
    =\;&\frac{\sqrt\pi}{2}\sqrt{\frac{2}{A}}erf\big(\sqrt{\frac{A}{2}}(\gamma-\frac{B}{A})\big)\Big|_\xi^\infty
    =\;\sqrt\frac{\pi}{2A}\Big(1 - erf\big(\sqrt{\frac{A}{2}}(\xi-\frac{B}{A})\big)\Big)
\end{align}
Finally, substitute the result of Eq.~\eqref{eq:errorfunc} to Eq.~\eqref{eq:integral}, and replace the integral in  Eq.~\eqref{eq:gauss} with the result of Eq.~\eqref{eq:integral}, will lead to the final expression of  Eq.~\eqref{eq:mulvarrank_ll}.
\subsection*{Gradients for the final objective function Eq.~\eqref{eq:mulvarrank_obj}}
Let us first denote the final objective function in Eq.~\eqref{eq:mulvarrank_obj} as $\mathcal L_{DMR}$, and the gradients $\frac{\partial \mathcal L_{DMR}}{\partial\Theta}$ with respect to each parameter in the parameter set $\Theta$ are given as follows:

\noindent\textbf{Gradients for $\bm U, \bm V, \bm W$:}
We can first take the derivatives with respect to the predicted pairwise difference vector $\hat{\bm d}_{u,i,j}$ obtained from the latent factors, and then take derivatives with respect to each latent factor matrix with the chain rule.
\begin{align*}
    \frac{\partial \mathcal L_{DMR}}{\partial\hat{\bm d}_{u,i,j}}
    =& -\frac{1}{2}(\bm\Sigma_{uij}^{-1}+\bm\Sigma_{uij}^{-\top})\hat{\bm d}_{u,i,j} + \frac{(\bm d_{u,i,j}\bm\Sigma_{uij}^{-1}\hat{\bm d}_{u,i,j}^\top)(\bm d_{u,i,j}\bm\Sigma_{uij}^{-1})}{\bm d_{u,i,j}\bm\Sigma_{uij}^{-1}\bm d_{u,i,j}^\top}\\
    &+ \frac{1}{1-erf(Z)}\cdot\frac{2}{\sqrt{\pi}}e^{-Z^2}\cdot\frac{\bm d_{u,i,j}\bm\Sigma_{uij}^{-1}}{\sqrt{2\bm d_{u,i,j}\bm\Sigma_{uij}^{-1}\bm d_{u,i,j}^\top}}
\end{align*}
where $Z=\sqrt{\frac{A}{2}}(\xi-\frac{B}{A})$. As $\hat{\bm d}_{u,i,j} = \big(U_u*(V_i-V_j)\big)W^\top$, with the chain rule of gradients, we have:
\begin{flalign}
\label{eq:grad_uvw}
    &\frac{\partial \mathcal L_{DMR}}{\partial\bm U_u} = \frac{\partial \mathcal L_{DMR}}{\partial\hat{\bm d}_{u,i,j}} \big((\bm V_i - \bm V_j)*\bm W\big) \\\nonumber
    &\frac{\partial \mathcal L_{DMR}}{\partial\bm V_i} = \frac{\partial \mathcal L_{DMR}}{\partial\hat{\bm d}_{u,i,j}} \big(\bm U_u*\bm W\big)  \\\nonumber
    &\frac{\partial \mathcal L_{DMR}}{\partial\bm V_j} = -\frac{\partial \mathcal L_{DMR}}{\partial\hat{\bm d}_{u,i,j}} \big(\bm U_u*\bm W\big)  \\\nonumber
    &\frac{\partial \mathcal L_{DMR}}{\partial\bm W} = (\frac{\partial \mathcal L_{DMR}}{\partial\hat{\bm d}_{u,i,j}})^\top \big(\bm U_u*(\bm V_i - \bm V_j)\big)
\end{flalign}
When the Hadmard product is applied to $\bm W$, it broadcasts to each row of $\bm W$. Note that the gradients for the priors on $\bm U, \bm V, \bm W$ are the same as standard $l_2$ regularization and are not included above.

\noindent\textbf{Gradients for $\bm\Sigma_u^U$ and $\bm\Sigma_i^V$:}
We first derive the gradients for the original covariance matrices $\bm\Sigma$ (including gradients on the normal-inverse-Wishart prior). Then we present the gradients on the decomposed matrices $L$ with the chain rule for practical optimization.
\begin{flalign}
    &\frac{\partial \mathcal L_{DMR}}{\partial\bm\Sigma_{uij}} 
    = -\frac{1}{2}\bm\Sigma_{uij}^{-\top} + \frac{1}{2}(\bm\Sigma_{uij}^{-\top}\hat{\bm d}_{u,i,j}\hat{\bm d}_{u,i,j}^\top\bm\Sigma_{uij}^{-\top}) &\\\nonumber
    &+\frac{1}{2A^2}\big[2AB(\bm\Sigma_{uij}^{-\top}\bm d_{u,i,j}\hat{\bm d}_{u,i,j}^\top\bm\Sigma_{uij}^{-\top}) + B^2(\bm\Sigma_{uij}^{-\top}\bm d_{u,i,j}\bm d_{u,i,j}^\top\bm\Sigma_{uij}^{-\top}) \big] &\\\nonumber
    &+\frac{\bm\Sigma_{uij}^{-\top}\bm d_{u,i,j}\bm d_{u,i,j}^\top\bm\Sigma_{uij}^{-\top}}{2A} - \frac{1}{1-erf(Z)}(\frac{2}{\sqrt{\pi}}e^{-Z^2})\cdot\frac{\partial Z}{\partial \bm\Sigma_{uij}}
\end{flalign}
where $\frac{\partial Z}{\partial\bm\Sigma_{uij}}$ is given by:
\begin{flalign}
    \nonumber&\frac{\partial Z}{\partial\bm\Sigma_{uij}} 
    = -\frac{\xi}{2\sqrt{2A}}(\bm\Sigma_{uij}^{-\top}\bm d_{u,i,j}\bm d_{u,i,j}^\top\bm\Sigma_{uij}^{-\top})\\\nonumber 
    &+ \frac{1}{2A}\Big[\sqrt{2A}(\bm\Sigma_{uij}^{-\top}\bm d_{u,i,j}\hat{\bm d}_{u,i,j}^\top\bm\Sigma_{uij}^{-\top}) - \frac{B}{2\sqrt{2A}}(\bm\Sigma_{uij}^{-\top}\bm d_{u,i,j}\bm d_{u,i,j}^\top\bm\Sigma_{uij}^{-\top})\Big]
\end{flalign}
Recall that $\bm\Sigma_{uij} = \bm\Sigma_{ui} + \bm\Sigma_{uj}$ and $\bm\Sigma_{ui} = \lambda\bm\Sigma_u^U + (1-\lambda)\bm\Sigma_i^V$. With the chain rule plus the gradients of the regularizations, we have:
\begin{equation*}
\begin{split}
    \frac{\partial \mathcal L_{DMR}}{\partial\bm\Sigma_{u}^U} &= 2\lambda \frac{\partial \mathcal L_{DMR}}{\partial\bm\Sigma_{uij}} + \frac{1}{2}({\bm\Sigma_{u}^U}^{-1}\Psi{\bm\Sigma_{u}^U}^{-1})^\top - \frac{\nu+K+1}{2}{\bm\Sigma_{u}^U}^{-\top}\\
    \frac{\partial \mathcal L_{DMR}}{\partial\bm\Sigma_{i}^V} &= (1-\lambda) \frac{\partial \mathcal L_{DMR}}{\partial\bm\Sigma_{uij}} + \frac{1}{2}({\bm\Sigma_{i}^V}^{-1}\Psi{\bm\Sigma_{i}^V}^{-1})^\top - \frac{\nu+K+1}{2}{\bm\Sigma_{i}^V}^{-\top} \\
    \frac{\partial \mathcal L_{DMR}}{\partial\bm\Sigma_{j}^V} &= (1-\lambda) \frac{\partial \mathcal L_{DMR}}{\partial\bm\Sigma_{uij}} + \frac{1}{2}({\bm\Sigma_{j}^V}^{-1}\Psi{\bm\Sigma_{j}^V}^{-1})^\top - \frac{\nu+K+1}{2}{\bm\Sigma_{j}^V}^{-\top}\\
\end{split}
\end{equation*}
Lastly, as each covariance matrix is decomposed as $\bm\Sigma = \bm L\bm L^\top$ for validity, we need to update $L$ during optimization:
\begin{equation*}
\begin{split}
    \frac{\partial \mathcal L_{DMR}}{\partial L_u^U} &= 2\frac{\partial l}{\partial \bm\Sigma_u^U}L_u^U,\,\,\,\,\frac{\partial \mathcal L_{DMR}}{\partial L_i^V} = 2\frac{\partial l}{\partial \bm\Sigma_i^V}L_i^V,\,\,\,\,\frac{\partial \mathcal L_{DMR}}{\partial L_j^V} = 2\frac{\partial l}{\partial \bm\Sigma_j^V}L_j^V
\end{split}
\end{equation*}

\end{document}